\documentclass[aastex]{article}
\usepackage[onecolumn]{emulateapj5}

\lefthead{Crittenden, Natarajan,  Pen \& Theuns}
\righthead{Discriminating weak lensing from intrinsic spin correlations }

\def\gs{\mathrel{\raise0.35ex\hbox{$\scriptstyle >$}\kern-0.6em 
les
\lower0.40ex\hbox{{$\scriptstyle \sim$}}}}
\def\ls{\mathrel{\raise0.35ex\hbox{$\scriptstyle <$}\kern-0.6em 
ggles
\lower0.40ex\hbox{{$\scriptstyle \sim$}}}}
\def\ltorder{
\mathrel{\raise.3ex\hbox{$<$}\mkern-14mu\lower0.6ex\hbox{$\sim$}}
}
\def\gtorder{
\mathrel{\raise.3ex\hbox{$>$}\mkern-14mu\lower0.6ex\hbox{$\sim$}}
}

\def\ep_s{\epsilon^{(s)}}

\def\ep{\epsilon}
\def\epp{\epsilon_{+}}
\def\epc{\epsilon_{\times}}
\def\xip{\xi_{+}}
\def\xic{\xi_{\times}}

\input psfig
\begin{document}
\title{Discriminating weak lensing from intrinsic spin correlations 
using the curl-gradient decomposition 
} 

\author{Robert G. Crittenden$^1$, Priyamvada Natarajan$^{2}$,
Ue-Li Pen$^3$
\& Tom Theuns$^4$}
\affil{1 Department of Applied Mathematics and Theoretical
Physics, Wilberforce Road, Cambridge CB3 0WA, UK} 
\affil{2 Department of Astronomy, Yale University, New Haven, CT 06511, USA}
\affil{3 CITA,
McLennan Labs, University of Toronto, Toronto, M5S 3H8}
\affil{4 Institute of Astronomy, Madingley Road, Cambridge 
CB3 0HA, UK}

\begin{abstract}
The distortion field defined by the ellipticities of galaxy shapes as
projected on the sky can be uniquely decomposed into a gradient and a
curl component. If the observed ellipticities are induced by weak 
gravitational lensing, 
then the distortion field is curl free.  Here we show that, in contrast, 
the distortion field resulting from 
intrinsic spin alignments
is not curl free. 
This provides a powerful discriminant
between lensing and intrinsic contributions to observed
ellipticity correlations. 
We also show how these contributions can be
disentangled statistically from the ellipticity correlations or 
computed locally from circular integrals of the ellipticity field.   
This allows for an unambiguous detection
of intrinsic galaxy alignments in the data.  When the distortions are 
dominated by lensing, as occurs at high redshifts,  
the decomposition provides a valuable tool for understanding properties of 
the noise and systematic errors. 
These techniques can be applied equally well to the polarization of the 
microwave background, where it can be used to separate curl-free scalar 
perturbations from those produced by gravity waves or defects.  
\end{abstract}

\section{Introduction}
The shapes of distant galaxies can be distorted due to gravitational
lensing by the intervening matter distribution (Gunn 1967; see
e.g. Bartelmann \& Schneider 1999 for a recent review and further
references). While the distortions are usually small, a signal
can still be detected statistically since neighboring objects will be
deformed in a similar way, thereby producing measurable correlations in
galaxy shapes. Recently, shape distortions of order 1\%  
have been detected by several groups in deep galaxy surveys 
on scales from one up to several arc minutes 
(van Waerbeke et al. 2000, Wittman
et al. 2000, Bacon, Refregier \& Ellis 2000, Kaiser, Wilson \&
Luppino 2000).  The amplitude of these distortions appears consistent with 
that predicted from weak lensing by large scale structure.  

Weak lensing is not the only possible source of galaxy shape correlations; 
these can also arise between physically close galaxies 
as a consequence of the galaxy formation process or subsequent
interactions. 
One possible mechanism for this is the coupling of galaxy spins: 
galaxy disks tend to be oriented perpendicular to their angular
momentum vectors, so angular momentum couplings of neighbors will
induce alignments in the projected galaxy shapes. This will also be
true but to a lesser extent for elliptical galaxies if they rotate
along their shortest axis.   The amplitude of the expected shape 
correlations from angular momentum couplings have recently been studied in 
numerical simulations by Heavens et al. (2000) and analytically by Crittenden 
et al. (CNPT 2000), while alternative mechanisms for intrinsic correlations 
have also been suggested (Croft and Metzler 2000; Catelan et al. 2000).  
Intrinsic correlations will be especially important in relatively shallow 
surveys such as the 2dF or the Sloan Digital Sky survey, 
where the median redshift is $\ll 1$. 
Evidence for the existence of such intrinsic correlations in the 
nearby universe has been recently presented by 
Pen, Lee \& Seljak (2000) and Brown et al. (2000), looking  
in the Tully and the Super COSMOS galaxy catalogs, respectively.

One way of disentangling the intrinsic shape correlations 
from those induced by weak lensing is to examine the 
patterns of the average galaxy shapes.  The shapes of galaxies are 
typically described by two degrees of freedom; their average ellipticity 
and orientation.  
The distribution of galaxy shapes can be described by a symmetric and
traceless 2D tensor field. In general, any such tensor field can be
written as a sum of two terms, one of which is curl-free and the other
is divergence-free. In analogy with the radiation field in 
electromagnetism, these
are usually referred to as the `electric' ($E$) and the
`magnetic' ($B$) component, respectively. 

Lensing by a point mass will create a tangential, curl-free distortion pattern. 
The most general distortion field produced by lensing will be 
a linear superposition of such patterns  
and will also be a curl-free, (i.e. $E$-type) field (Kaiser
1992, Stebbins 1996).  
However, as we show below, the distortion field resulting from intrinsic spin 
alignments has $E$ and $B$-type modes of the same order of magnitude. 
This property will enable us 
to uniquely disentangle angular
momentum correlations and to subtract their contribution from
a measured distortion field in order to more accurately compute 
and isolate the distortions
induced by lensing alone. 
When the deformations are dominated by lensing, the $E-B$ 
decomposition can improve 
the signal to noise level, since noise and other 
systematic effects are expected to contribute to both the $E$ and $B$ channels. 
Current surveys have measured the sum of $E$ and $B$ powers, thus doubling the 
noise power relative to the decomposition strategy that we propose here. 

The expected amplitude of galaxy shape correlations is fairly small.
In order to measure it given the large scatter in the 
intrinsic shapes of galaxies, many galaxies must be observed.  At present, 
observations of mean ellipticities are dominated by the intrinsic scatter. 
Direct decomposition into $E$ and $B$ modes is a non-local operation, requiring 
derivatives of these noisy observations and so is very problematic.  
Here we show how the correlation functions of the observable 
ellipticities can be directly 
converted into correlations of the $E$ and $B$-modes.  

It is also useful to have locally defined quantities which reflect the 
$E$ and $B$ decomposition.  Kaiser et al. (1994) and Schneider et al. (1998) 
looked at this issue in the 
context of lensing and developed a statistic known as the `aperture mass' 
which enabled them to put a lower bound on the projected mass in a localized 
region of the sky.  More generally, this statistic gives a direct, local  
measure of the electric contribution to the distortion field, and a similar 
observable can be evaluated to measure the magnetic component.  We develop this 
formalism here and relate the correlations of these local $E$ and $B$ 
estimators to 
those of the observable ellipticities.   

This paper is organized as follows. We begin by defining the $E-B$
decomposition in terms of observed ellipticities, and demonstrate that
the lensing distortions are curl free, whereas those due to intrinsic
alignments are not. In Section~3, we discuss estimators of the $E-B$
correlations and their relation to correlators of the ellipticities. 
In Section~4, we define the local $E$ and $B$ measures and calculate their 
correlations. 
We conclude in sections 5-6 with a cookbook style summary on
how the $E-B$ decomposition can be derived from the statistical weak
lensing surveys.

\section{Decomposition of the distortion field}

The projected shape of a galaxy on the sky can be approximated by an
ellipse with semi-axes $a$ and $b$ ($a > b$), of which the major axis
makes an angle $\psi$ with respect to the $x$-axis of the chosen
coordinate system. It can then be concisely written as a complex
number,
\begin{eqnarray}
\epsilon&=&{(a^2 -
b^2)\over (a^2 + b^2)} e^{2i\psi}
=\epsilon_{+} + i \epsilon_{\times}\,,
\end{eqnarray}
where $\epsilon_{+}=|\epsilon|\cos(2\psi)$ and
$\epsilon_{\times}=|\epsilon|\sin (2\psi)$. The phase dependence
$\propto e^{2i\psi}$ expresses the fact that the ellipticity is
invariant under a rotation over $\pi$ radians.  Note that the two
components $\epsilon_{\times}$ and $\epsilon_{+}$ are analogous to the
$Q$ and $U$ Stokes parameters for linearly polarized light.  
Given a distribution of galaxies on the sky with measured
ellipticities, the complex scalar ellipticity field defines a
traceless, symmetric $2 \times 2$ tensor field,
\begin{eqnarray}
[\gamma]_{ab} = \left[\begin{array}{cc}\epsilon_{+} & \epsilon_{\times}\\ 
\epsilon_{\times} &-\epsilon_{+} \end{array}\right]\,.
\label{eq:gamma}
\end{eqnarray}
We have approximated the sky as flat and followed the derivation of
Kamionkowski et al. (1998). See Stebbins (1996) for the generalization
to a curved sky.

The shear field Eq.~(\ref{eq:gamma}) can be written in terms of a
gradient or \lq $E$\rq~-piece and a curl or pseudo-scalar \lq
$B$\rq~-piece (Stebbins 1996) by introducing two scalar functions
$\Phi_E$ and $\Phi_B$,
\begin{eqnarray}
\gamma_{ab}({\mathbf x})\,=\,(\partial_{a} \partial_{b} - 
\frac{1}{2} \delta_{ab} \nabla^2)\,\Phi_{E}({\mathbf x}) + 
\frac{1}{2} (\epsilon_{cb} \partial_{a} \partial_{c} +
\epsilon_{ca} \partial_{c} \partial_{b})\,\Phi_{B}
({\mathbf x}),
\label{eq:eb1}
\end{eqnarray}
where $\epsilon_{ab}$ is the anti-symmetric tensor.
Each component of the ellipticity field can be written as a 
function of $\Phi_E$ and $\Phi_B$ as
\begin{eqnarray}
\epsilon_{+} &=& \gamma_{xx} = - \gamma_{yy} = \frac{1}{2}(\partial_{x}
\partial_{x} - \partial_{y} \partial_{y})\,\Phi_{E}
({\mathbf x}) - \partial_{x} \partial_{y}\,\Phi_{B}
({\mathbf x}) \nonumber \\ 
\epsilon_{\times} &=& \gamma_{yx} = \gamma_{xy} = \partial_{x}
\partial_{y} \Phi_{E} ({\mathbf x}) + \frac{1}{2} (\partial_{x}
\partial_{x} - \partial_{y} \partial_{y})\,\Phi_{B}
({\mathbf x}).
\label{eq:defepEB}
\end{eqnarray}
The $E$ and $B$ parts can be extracted explicitly from the shear tensor
by applying the $\nabla^4$ operator,
\begin{eqnarray}
{\nabla^4} \Phi_E = 2\,\partial_{a} 
\partial_{b}\gamma_{ab}; \,\,\,
{\nabla^4} \Phi_B = 2\,\epsilon_{ab}\,\partial_{a}
\partial_{c}\gamma_{bc}.
\end{eqnarray}
The relation between the functions $\Phi_E$ and $\Phi_B$ and the projected
gravitational potential will become obvious in what follows.
It is useful to perform the $E-B$ decomposition in terms of variables
that have the same dimension as the measured ellipticities
(Kamionkowski et al. 1998): $\gamma_E \equiv {1 \over 2} \nabla^2
\Phi_E$ and $\gamma_B \equiv {1 \over 2} \nabla^2 \Phi_B$. These are
related to the ellipticities by
\begin{eqnarray}
\nabla^2 \gamma_E = \partial_{a}\partial_{b}\gamma_{ab} & =&  
(\partial_{x} \partial_{x} - \partial_{y} \partial_{y}) \epp +  
2 \partial_{x} \partial_{y} \epc \nonumber \\
\nabla^2 \gamma_B = \epsilon_{ab}\,\partial_{a} \partial_{c}\gamma_{bc} &=& 
(\partial_{x} \partial_{x} - \partial_{y} \partial_{y}) \epc -  
2 \partial_{x} \partial_{y} \epp\,.
\label{eq:g_eb}
\end{eqnarray}

Since only the {\em derivatives} of $\gamma_E$ and $\gamma_B$ are
defined in terms of the ellipticity field, $\gamma_E({\mathbf{x}})$ and
$\gamma_B({\mathbf{x}})$ are ambiguous up to a constant and linear
gradient term. In the same way, constant and linear gradient terms in
the measured ellipticities should not impact the $E-B$ decomposition.
To see this, consider an ellipticity field where $\epp({\mathbf{x}}) =
x$ and $\epc({\mathbf{x}}) = 0.$ This can either be a consequence of a
pure $E$-mode ($\Phi_E = x^3/3; \Phi_B =0$), the result of a pure
$B$-mode ($\Phi_E = 0; \Phi_B = x^2 y/2 + y^3/6$), or a linear
combination of the two.

A rotation of the basis axes translates $\epp$ into $\epc$ and vice
versa, but does not affect the $E-B$ decomposition.  In particular, the
ellipticity measured in a basis which is at an angle $\varphi$ relative
to the original basis is given by:
\begin{equation}
\epsilon_{+}'= \epsilon_{+} \cos 2\varphi - \epsilon_{\times} \sin
2\varphi\,;\,\,\, 
\epsilon_{\times}'= \epsilon_{+} \sin 2\varphi + \epsilon_{\times} \cos 2\varphi.
\end{equation}
Thus, a global rotation of $\pi/4$ transforms $\epp' =  -\epc;
\epc' =  \epp$, but since the position vectors are also rotated, the $E-B$ 
decomposition remains invariant.   
However, one can also take an ellipticity field and rotate each ellipticity   
individually by $\pi/4$, keeping the position vectors fixed. 
This new ellipticity map has the $E$ and $B$ modes of the original map 
interchanged: 
$\gamma_E' = - \gamma_B ; \gamma_B' = \gamma_E$.  
 
\subsection{Distortions due to lensing}

In the case of gravitational lensing, the resultant distortion field
$\gamma$ can be written in terms of a gravitational deflection
potential $\psi$ as (e.g. Bartelmann and Schneider 1999)
\begin{eqnarray} 
\gamma_{ab}({\mathbf{x}}) = 
(\partial_a \partial_b - {1 \over 2} \delta_{ab} \nabla^2 ) 
\psi({\mathbf{x}})\,.
\label{eq:Gshear}
\end{eqnarray}
The deflection potential $\psi$ is a convolution over the projected
surface mass density $\kappa({\mathbf{x}})$, $\psi({\mathbf{x}}) = {1 \over
\pi} \int d {\mathbf{x}}' \kappa({\mathbf{x}}') \ln|{\mathbf{x}} -
{\mathbf{x}}'|$. Comparing this expression to the $E-B$ decomposition
of Eq.~(\ref{eq:eb1}) we can identify $\Phi_E({\mathbf{x}}) =
\psi({\mathbf{x}})$ and $\Phi_B = 0.$ Thus for the shear field induced
by lensing, the $E$-mode is related to $\kappa$, the projected surface
mass density in units of the critical surface mass density for a given
configuration of source and lens, and the $B$-mode is identically
zero (as was discussed by Kaiser 1995; Kamionkowski et al. 1998).
Note that weak lensing only approximately gives pure $E$-modes, as $B$-modes
may arise when the light is bent in more than one scattering event.  
However, these $B$-modes arise at higher order and are suppressed relative to 
the $E$-modes. 

If one measures the $E$-mode contribution of a given map and then
rotates every measured ellipticity by $\pi/4$ and repeats the same measurement,
one obtains an estimate of the $B$-contribution which should be
consistent with zero for a pure lensing signal. Therefore, the absence
of $B$-modes naturally provides a robust test for isolating the lensing 
component of the distortion field and provides an estimate of the 
noise level of the data (Kaiser 1992).

Note that lensing is not the only possible source of curl-free correlations. 
If the shapes of galaxies are primarily determined by tidal stretching, this 
would lead to intrinsic correlations (Catelan, Kamionkowski \& Blandford, 2000; 
Croft \& Metzler, 2000).  For these shape distortions, the observed 
ellipticities 
are also linear in the tidal field, leading to pure electric modes just 
as in lensing.  For spiral galaxies which have had many dynamical 
times to evolve, tidal stretching is likely to be small compared to the 
contribution from spin alignments. 
Even for elliptical galaxies, a bulk rotation of as small as 1 km/sec
would erase the galaxy's original alignment.  Almost
all observed ellipticals rotate faster than that, so the shape-shear
alignment in Catelan et al. (2000) is unlikely to be observable.    
However, it could be significant for larger 
objects like clusters which are dynamically much younger. 

\subsection{Distortions due to angular momenta alignments}

Shape correlations between galaxies can also arise from alignments in
the direction of their angular momenta. While this is particularly
true for spiral galaxies, given the assumption that their disks are
perpendicular to the angular momentum vectors, it is true to a lesser
extent for elliptical galaxies as well. 
Ellipticals probably rotate
about their intermediate axis (Dubinski 1992), so averaging over all
statistical randomized alignments, the average major axis is
perpendicular to the angular momentum vector, just like a spiral galaxy.
The strength of the
correlation signal has recently been studied in numerical simulations
by Heavens et al. (2000).  
We have recently attempted to model these theoretically (CNPT 2000) 
by assuming angular momentum is induced by tidal torques.  
Following the formalism developed by Catelan and Theuns (1996), the correlations
can be calculated for Gaussian initial fluctuations using linear theory 
coupled with the Zeldovich approximation.  

The induced intrinsic correlations of ellipticities will primarily
result from correlations in the direction of the angular momenta. As
discussed in CNPT this implies that the ellipticities are effectively
quadratic in the angular momenta,
\begin{eqnarray}
\bar{\ep}_+  \propto {1 \over 2}
(\hat{T}_{x i}\hat{T}_{i x} - \hat{T}_{y i}\hat{T}_{i y});
\, \, \bar{\ep}_{\times} \propto
 {1 \over 2}
\hat{T}_{x i}\hat{T}_{i y}.
\label{eq:Tshear}
\end{eqnarray}
Here, $T_{ij}\propto \partial_i\partial_j\phi$ is the shear of the
gravitational potential, and $\hat{T}$ denotes the shear tensor normalized by 
$\sqrt{T_{ij}T_{ij}}$. Note that $i$ and $j$ run over three coordinates, $x, y, z$ 
in contrast to above where two dimensional (projected) quantities were considered.  
The
quadratic dependence on the shear in Eq.~(\ref{eq:Tshear}) is
fundamentally different from the linear one appropriate for the lensing
case, Eq.~(\ref{eq:Gshear}), and as a result the $B$-modes are
non-zero.

In particular,  it is straight forward to show that
\begin{eqnarray}
\nabla^2 \gamma_E \propto \hat{T}_{x i, x x} \hat{T}_{i x} + 
\hat{T}_{y i, y y} \hat{T}_{i y} + 
\hat{T}_{x i, x y} \hat{T}_{i y} +
\hat{T}_{y i, x y} \hat{T}_{i x} +
(\hat{T}_{x i, x} + \hat{T}_{y i, y})^2 . 
\end{eqnarray}
and similarly 
\begin{eqnarray}
\nabla^2 \gamma_B \propto \hat{T}_{x i, x x} \hat{T}_{i y}  - 
\hat{T}_{y i, y y} \hat{T}_{i x} - 
\hat{T}_{x i, x y} \hat{T}_{i x} + 
\hat{T}_{y i, x y} \hat{T}_{i y}. 
\end{eqnarray}
The amplitude of the $B$-modes is comparable to that of the $E$-modes, 
as is shown in Figure 1. 
The presence of $B$-modes provides a mechanism for disentangling the
correlations which arise from lensing from those resulting from
intrinsic alignments due to angular momentum couplings. 

This is not the sole means of disentangling intrinsic correlations from 
weak lensing.  
The observed ellipticity correlations from intrinsic alignments 
are strongest at low redshifts, while the lensing signal is larger when the 
sources are at higher redshifts. 
In addition, morphology distinctions will be useful, since 
intrinsic correlations of elliptical galaxies are smaller 
than those of spirals because they are intrinsically more round.  
Further possible methods for distinguishing lensing from intrinsic 
correlations are discussed in Catelan, Kamionkowski and Blandford (2000) 
and CNPT (2000).   

\section{Correlation estimators} 
The decomposition into curl and gradient contributions is most straight 
forwardly performed in Fourier space (Kamion-kowski et al. 1998).  
However, often it is useful to consider this in real space, as many issues 
which might complicate matters in Fourier space, such as finite field size or  
patchy sampling, are more easily handled in real space.  In this section 
we deal with performing the decomposition statistically in real space 
using the ellipticity two point functions.  This approach is particularly 
relevant when the observations are noise dominated, such as is the 
case when there are relatively few galaxies with which to measure 
the mean ellipticity, 
which is the case for the current surveys on small scales where most
of the lensing signal lies.
In the next section, we will address the issue of a local decomposition.  

\subsection{Correlations in $\gamma_E$ and $\gamma_B$}

Here we will relate the correlations of the electric and magnetic shear 
directly to correlations of the ellipticity.  
The real space correlation function of $\gamma_E$ is 
related to correlation of its corresponding potential, $\Phi_E$, by 
\begin{eqnarray}
\xi_E \equiv 
\langle \gamma_E ({\mathbf{x}}) \gamma_E ({\mathbf{x+r}})  \rangle 
= {1 \over 4} \langle \nabla^2 \Phi_E({\mathbf{x}}) \nabla^2 
\Phi_E({\mathbf{x+r}}) \rangle  = {1 \over 4} \nabla^4 \Xi_E(r),
\label{eq:g_e}
\end{eqnarray}
where $\Xi_E(r) \equiv 
\langle\Phi_E({\mathbf{x}})\Phi_E({\mathbf{x+r}})\rangle$.  
An analogous relation holds for $\gamma_B$ correlations, while the cross 
correlation, $\langle \gamma_E \gamma_B \rangle$, is zero if the field is 
invariant under parity transformations.

These correlations can be directly computed from the observed
correlations in $\epp$ and $\epc$, defined through
\begin{equation}
C_1(r,\varphi) \equiv \langle \epsilon_{+}({\mathbf{x}}) \epsilon_{+} 
({\mathbf{x+r}})\rangle \,;\,\,
C_2(r,\varphi) \equiv \langle \epsilon_{\times}({\mathbf{x}}) 
\epsilon_{\times} ({\mathbf{x+r}})
\rangle 
\end{equation}
where the ensemble average is over pairs with separation $r$, for which
the separation vector $\mathbf{r}$ makes an angle $\varphi$ with
respect to the chosen basis. The sum of these correlations is
rotationally invariant, but their difference depends explicitly on the
choice of orientation of the coordinate axes (Kamionkowski et al. 1998).

The required relation between the observed correlations $C_1$ and
$C_2$, and the $E-B$-correlations $\xi_E$ and $\xi_B$ follows from
their definitions using Eq.~(\ref{eq:defepEB}) written in terms of
derivatives with respect to the separation $r$:
\begin{eqnarray}
C_1(r, \varphi) &=& {1 \over 8} \nabla^4 [\Xi_{E}(r) + \Xi_{B}(r)] +
{1 \over 8} \chi 
[\Xi_{E}(r) - \Xi_{B}(r)] \cos 4\varphi \nonumber \\ 
C_2(r, \varphi) &=& {1 \over 8} \nabla^4 [\Xi_{E}(r) + \Xi_{B}(r)] - 
{1 \over 8} \chi 
[\Xi_{E}(r) - \Xi_{B}(r)]\cos 4\varphi ,
\label{eq:c1c2veb}
\end{eqnarray}
where $\nabla^4 = 8D^2 + 8r^2 D^3 + r^4 D^4$, the operator $\chi =
r^4 D^4$ and $D \equiv {1 \over r}{\partial \over \partial r}$.  
We wish to invert these equations to find expressions for $\xi_E$ and 
$\xi_B$ in terms of the observable correlation functions. 

This inversion  
can be done most easily in terms of basis independent 
correlation functions which we shall denote $\xi_+$ and
$\xi_\times$.  
Physically, $\xi_+$ corresponds to the correlation function
$\langle\epp({\mathbf{x}})\epp({\mathbf{x+r}})\rangle$ computed in such
a way that, for each pair of galaxies, one coordinate axis is always
taken parallel to the separation vector $\mathbf{r}$.  
A similar definition holds for
$\xi_\times$ in terms of $\langle\epc\epc\rangle$, while by isotropy, the 
expectation of the cross correlation is zero.  
These correlations are related by a rotation to $C_1(r,\varphi)$ and 
$C_2(r,\varphi)$ and satisfy the relations, 
\begin{equation}
\xi_+(r) +\xi_\times(r) = C_1(r,\varphi) +C_2(r,\varphi)\,;\,\,\,
[\xi_+(r) -\xi_\times(r) ]\cos(4\varphi)=C_1(r,\varphi) -C_2(r,\varphi)\,.
\label{eq:xipxix}
\end{equation}
In terms of these
new observables, Eq.~(\ref{eq:c1c2veb}) simplifies to
\begin{equation}	
4[\xi_{+}(r) + \xi_{\times}(r)] =  \nabla^4  [\Xi_E(r) + \Xi_B(r)] 
;\,\,\,
4[\xi_{+}(r) - \xi_{\times}(r)] =  \chi [\Xi_E(r) - \Xi_B(r)]. 
\end{equation}

Initial measurements of the lensing signal have focused primarily on
the variance of the magnitude of the ellipticity averaged over regions 
of a given size,   
and its fall off as the size of the regions is increased.
The variance is simply the value of the correlation $C_1 + C_2$,  
convolved with the appropriate window function, at zero lag. 
The variance has the advantage that it is a local quantity and is 
straight forward to measure.  However, the measurements of the 
variance at different scales have strongly correlated errors. 
In addition, since $C_1 + C_2 = {1 \over 8} 
\nabla^4(\Xi_E+\Xi_B)$, such measurements are unable to distinguish
$E$-modes from $B$-modes.  Therefore, it is advantageous to investigate 
the full correlation functions. 

The inversion of Eq.~(\ref{eq:c1c2veb}) can now be written as,
\begin{eqnarray}
\xi_E(r)  &= &{1 \over 2} [\xi_{+}(r) + \xi_{\times}(r)] 
+ {1 \over 2}\nabla^4 \chi^{-1} [\xi_{+}(r) - \xi_{\times}(r)] \label{eq:invE} \nonumber\\ 
\xi_B(r) &= & {1 \over 2}[\xi_{+}(r) + \xi_{\times}(r)] 
- {1 \over 2}\nabla^4 \chi^{-1} [\xi_{+}(r) - \xi_{\times}(r)]\label{eq:invB}\,.
\end{eqnarray}
An equivalent set of equations follows by applying $\chi
\,\nabla^{-4}$ operator to both sides of these equations.  
\begin{eqnarray}
\chi \,\nabla^{-4} \xi_E(r) = {1 \over 2} 
\chi \,\nabla^{-4}[\xi_{+}(r) + \xi_{\times}(r)] 
+ {1 \over 2}[\xi_{+}(r) - \xi_{\times}(r)]  \nonumber\\ 
\chi \,\nabla^{-4} \xi_B(r) = {1 \over 2}
\chi \,\nabla^{-4} [\xi_{+}(r) + \xi_{\times}(r)] 
- {1 \over 2}[\xi_{+}(r) - \xi_{\times}(r)]\,.
\end{eqnarray}
This form actually proves more useful in practice 
since, as we show below,  
$\chi\,\nabla^{-4}$ is more local than $\nabla^4 \chi^{-1}$. 
Note that these expressions assume statistical isotropy, that is, 
$\langle \epp \epc \rangle = 0$ in the basis where the axes are aligned 
with the galaxy separation vector. 

\subsection{Evaluation of $\nabla^4\,\chi^{-1}$ and $\chi\,\nabla^{-4}$} 
It is quite useful to take these relationships into Fourier space, and 
the operators $\nabla^4$ and $\chi$ have particularly simple
expressions when applied to the Bessel functions which arise in 
a Fourier transform. In particular,
\begin{eqnarray}
\nabla^4 J_0(kr) = k^4 J_0(kr);\,\,\,  \chi J_0(kr) = k^4
J_4(kr)\,.
\label{eq:j0j4}
\end{eqnarray}
When combined with the above equations, we can relate the correlation
functions directly to the $E$ and $B$ power spectra, reproducing the
relations Eqs.~(14) and (15) of Kamionkowski et al. (1998).
Also, it is straightforward to show that since the power spectrum 
is the Fourier transform of the correlation function, 
$\xi_E(r) = \int k dk J_0(kr) P_E(k)$ 
implies that  $\chi \,\nabla^{-4} \xi_E(r)
= \int k dk J_4(kr) P_E(k)$. 

The operators $\nabla^4\,\chi^{-1}$
and $\chi\,\nabla^{-4}$ are also most easily evaluated in Fourier space
and can be shown to take the
form,
\begin{eqnarray}
\nabla^4\,\chi^{-1}\,g(r) &=& \int {k dk\over 2\pi} J_0(kr) \int r' dr' J_4(kr') g(r') 
= \int r' dr' g(r') {\cal{G}}(r,r') \nonumber\\
\chi \nabla^{-4}\,g(r)    &=& \int {k dk\over 2\pi} J_4(kr) \int r' dr' J_0(kr') g(r') 
= \int r' dr' g(r') {\cal{G}}(r',r)\,,
\end{eqnarray}
where 
\begin{eqnarray}
{\cal{G}}(r',r) = \int {k dk\over 2\pi} J_0(kr) J_4(kr')\,.
\label{eq:G}
\end{eqnarray}
Note that since the $k$ integral is from zero to infinity, $r
r'{\cal{G}}(r,r')$ depends only on the ratio $r/r'$. This operator is not
simply a convolution as would be the case if ${\cal{G}}(r,r')$ 
were a function only
of the difference $|\mathbf{r -r'}|$.  In fact, the Fourier space operator takes
nearly the same form as the real space operator.  That is, if the
Fourier transform of $g(r)$ is $g(k)$, then the transform of $\int r'
dr' g(r') {\cal{G}}(r,r')$ is $\int k' dk' g(k') {\cal{G}}(k',k)$.

Alternatively, these operators can be written in integral form,
\begin{eqnarray}
\nabla^4\,\chi^{-1}\,g(r)\,&=&\,g(r)\,+\,4 \int_r^{\infty}
dr'\, \frac{g(r')}{r'} - 12 r^2 \int_r^{\infty}dr'\,\frac{g(r')}{r'^3} 
\nonumber \\ 
\chi\,\nabla^{-4}\,g(r)\,&=&\,g(r)\,+\,\frac{4}{r^2}\int_{0}^{r}\,dr'\,
g(r')\,r'\,-\,\frac{12}{r^4}\,\int_{0}^{r}\,dr'\,g(r')\,r'^3\,.
\end{eqnarray}
The second form is the more useful in practice, being more local.  To
evaluate it at some radius $R$, one only needs to know the form of
$g(r)$ for $r<R$, whereas the $\nabla^4\,\chi^{-1}$ requires
knowing $g(r)$ for $r> R$.

The particular form of the first of the integral operators follows from
choosing integration constants such that $\nabla^4\,(\Xi_{E}(r)
+\Xi_{B}(r))$ does not diverge at large separations. To obtain the
second form, we demand that $\chi (\Xi_{E}(r) - \Xi_{B}(r))$ and its
derivatives are well-behaved as $r \rightarrow 0$.  The origin of these
integration constants is the constant and linear gradient ambiguities
in the definitions of $\gamma_E$ and $\gamma_B$ in Section~2.

\subsection{Power law solutions} 

As an illustration, consider the case where the correlation functions
both behave as power laws with the same index, $\xi_+(r) = Ar^n $ and
$\xi_{\times}(r) = Br^n $.  Note that $\nabla^4 \chi^{-1} r^n =
f(n) r^n $, where $f(n) = (n^2 + 6n +8)/(n^2-2n)$. It then follows that
\begin{eqnarray}
\xi_{E,B} = f(n) \chi \nabla^{-4}\xi_{E,B}\, = {1 \over 2} 
[(A+B) \pm (A-B)f(n)]r^n.
\end{eqnarray} 
Note that for certain power laws, $n = -2$ or $n = -4$, 
the operator $\nabla^4 \chi^{-1} r^n = 0$, while it diverges 
when $n = 0$ or $n=2$.  The opposite is true for the inverse operator 
$\chi\nabla^{-4}$, which diverges when  $n = -2$ or $n = -4$ and is 
zero for $n = 0$ or $n=2$. 

In the case when the $B$-modes are exactly zero, as occurs
for gravitational lensing, one obtains $(A + B) = (A-B) f(n).$ It
follows that the ratio of the correlation functions is given by
\begin{eqnarray}
\xi_+/\xi_{\times} = \frac{n^2 + 2 n + 4} {4 (n+1)}. 
\end{eqnarray} 
This expression diverges when $n = -1$ which therefore implies that
$\xi_{\times} = 0$ in this case.  Kaiser (1992) also considered power
law spectra in the lensing case and presented results for a number of
spectral indices.  Our results agree qualitatively, but the agreement
is not exact.  Kaiser calculated these numerically and this may be the source
of the discrepancies.    

\subsection{Application to spin correlations}  

\begin{figure}[t] 
\centerline{\psfig{file=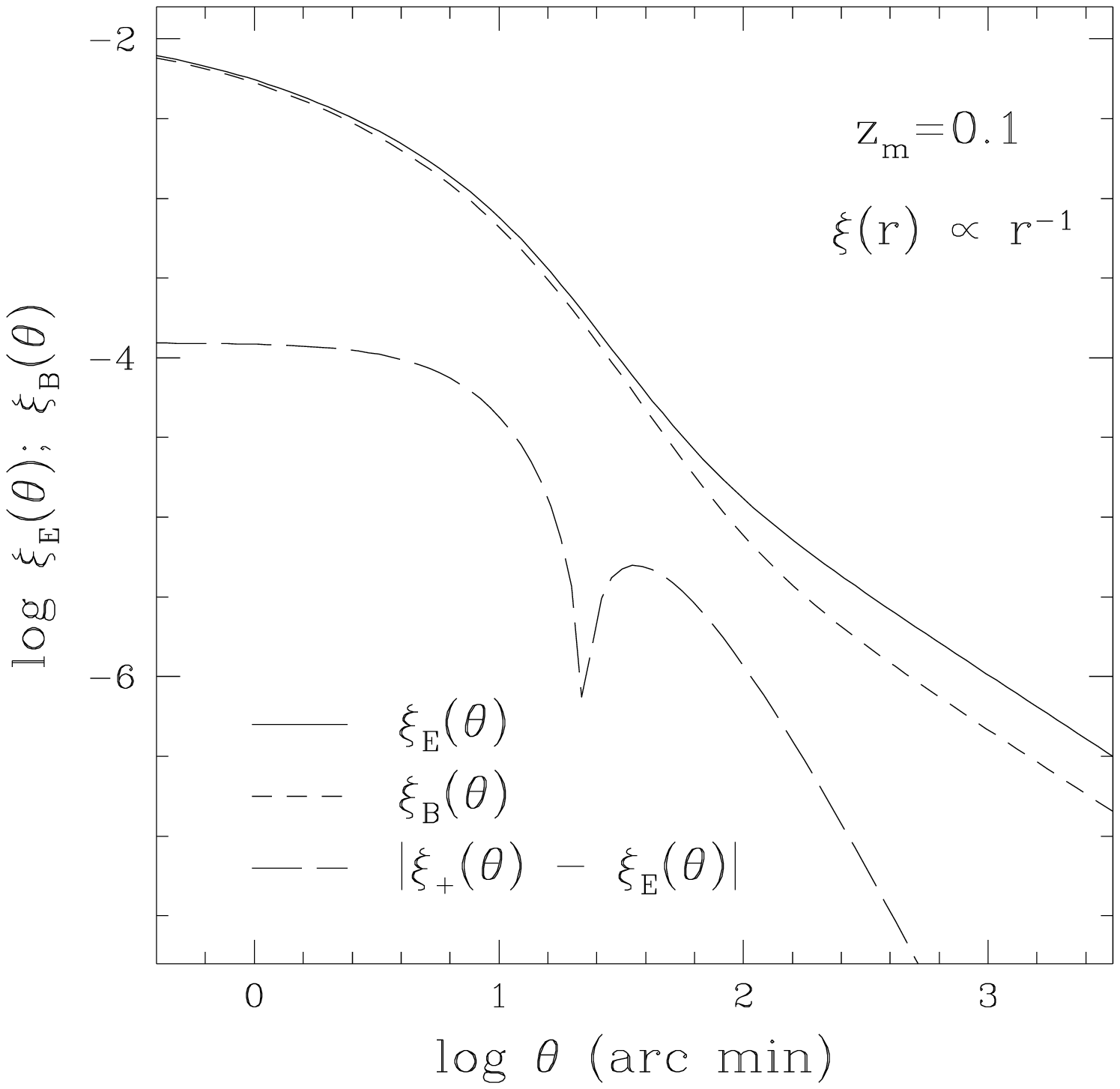,height=3.3in,width=3.3in}
\psfig{file=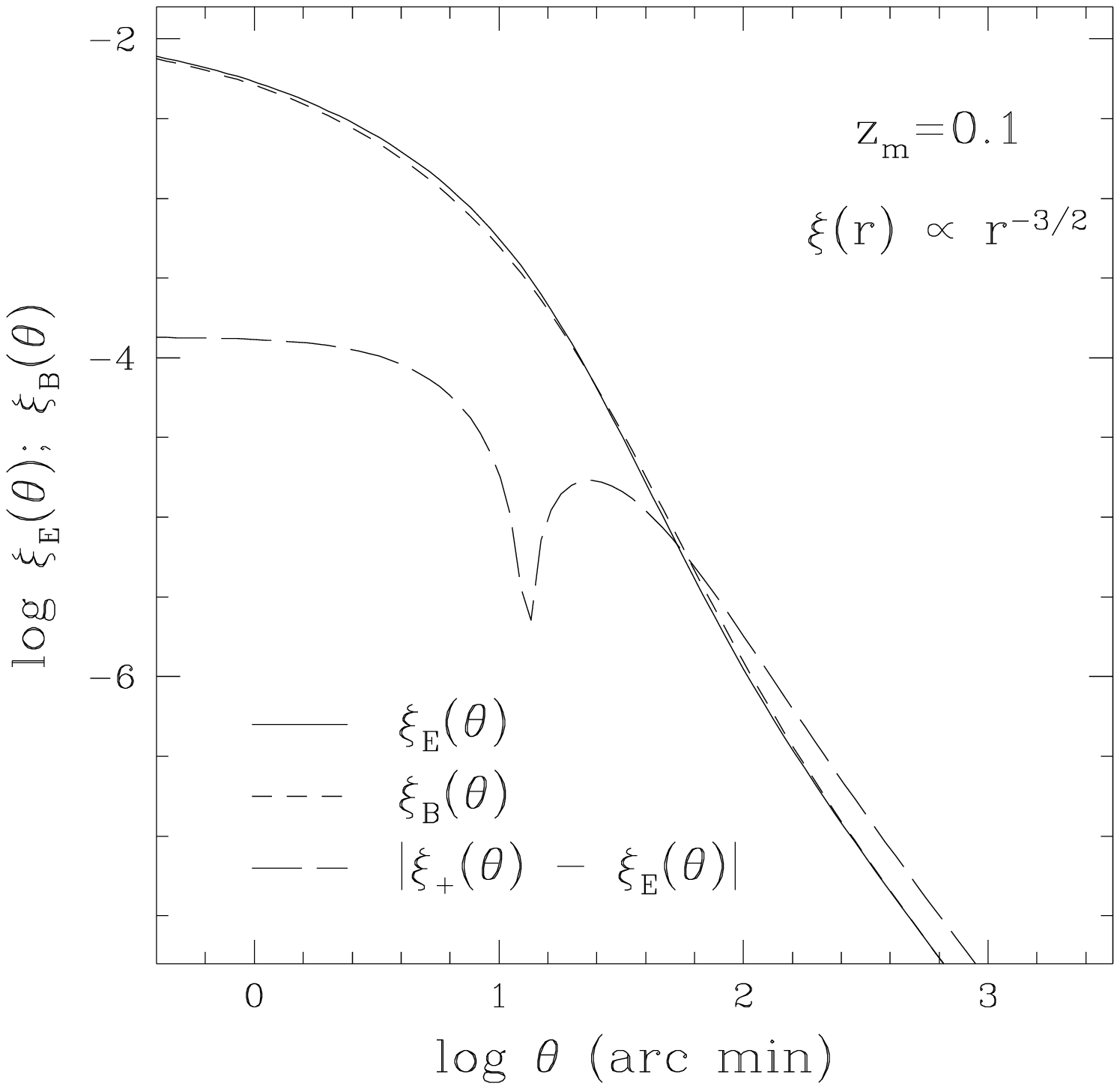,height=3.3in,width=3.3in}}
\caption{The $E$ and $B$-mode correlation functions for intrinsic 
spin correlations in the model of CNPT (2000).  The amplitude 
of the correlations are determined by the parameters $a$ and 
$\alpha$, which have been taken to be unity for simplicity.  
The mean redshift of the sources was taken to be $z_m = 0.1$ and 
the density correlations were taken to fall off as $r^{-1}$ in the 
left figure and as $r^{-3/2}$ in the right figure.  
Also plotted are the differences between $\xi_E$ and $\xi_+$, which is 
the same as the differences between $\xi_B$ and $\xi_\times$.  
In the left panel, the projected ellipticity correlations fall 
off as $\theta^{-1}$
and $\xi_E$ and $\xi_+$ are very nearly the same. 
This is not the case in general, as can be seen 
in the right panel where the projected correlations 
fall as $\theta^{-2}$.  However, the $\theta^{-2}$ is also special in that 
$\xi_E$ and $\xi_B$ are nearly identical.    
}
\label{fig-ebcorr}
\end{figure}

We can apply this technique to the correlations arising from intrinsic 
spin couplings.  The ellipticity correlations were calculated for the model
described in CNPT (2000), and we will not go into further detail here.  
Using the CNPT results and the above expressions, we can calculate the 
$E$ and $B$ correlation functions and these are presented in Figure 1.  
In contrast to the gravitational lensing cases, the $E$ and $B$ modes are seen 
to be of comparable magnitude in this model.  

Various parameter choices have been made in the examples we have shown, but  
we believe the $E-B$ decomposition to be largely independent of most of these.  
For simplicity, the figures assume the galaxies are effectively perfect 
disks ($a=1$ in the notation of CNPT).  
Using more realistic galaxy shapes will only suppress the 
overall amplitude of the correlations.  We have also assumed that the 
angular momentum directions correspond with those that would be predicted 
by linear theory ($\alpha = 1.$)  Non-linear evolution may affect the direction
of a galaxy's angular momentum, but as long as these changes are not coherent  
then they will only suppress the overall correlation amplitudes. 
Finally, we have assumed a mean redshift for the sources of $z_m = 0.1$. 
Changing this will change the angular scale at which a given level of 
correlations are seen, but it should not affect the nature of 
the $E-B$ decomposition.  

As discussed above, one factor which could affect the $E-B$ decomposition is 
the rate at which ellipticity correlations drop off.  
The case when the density correlation falls off as $r^{-1}$ implies that 
for large separations, the projected ellipticities fall off as $\theta^{-1}$, 
where $\theta$ is the angular separation. 
Since $f(-1) = 1$, $\xi_E$ will be the same as $\xi_+$, and $\xi_B$ will be 
the same as $\xi_\times$.  This is shown in the left panel of Figure 1. 
Also shown in the right panel is the case when the density correlation 
falls off as $r^{-3/2}$ and the projected ellipticity drops as $\theta^{-2}$. 
In this case, $f(-2) = 0$ and $\xi_E = \xi_B = {1 \over 2} 
(\xi_+ + \xi_\times).$  Note that here the $\times$ modes are actually 
anti-correlated at large separations. 

\section{Local correlators}
Here we will consider local estimators of the $E$ and $B$ modes.  
These are generalizations of the `mass aperture' formalism of 
the pure lensing case, 
proposed by Kaiser et al. (1994) and developed further by Schneider et al. 
(1998) and applied to CMB polarization by Seljak and Zaldarriaga (1998).
They showed that a convolution of the tangential shear with a 
given wavelet provided a measure of the projected mass convolved with a related
wavelet. More generally, the integrals of the tangential shape 
distortions can be directly 
related to the local 'electric' distortion.  Thus $E$-modes are associated 
with either tangential or radial patterns, as shown in Figure 2 (left panel.) 
$B$-modes are also related to the circular distortion pattern, but these have 
an associated `handedness' or orientation, as shown in the right panel of 
Figure 2.  

The local correlators are most easily defined by considering polar 
coordinates about a given point.  The $E$-mode is related to the 
tangential shear,
which is the local $\epp$ (or $Q$ Stokes parameter) in a coordinate system 
defined by the radial vector to that point.
The $B$-mode corresponds analogously with the local $\epc$ mode (or $U$ 
Stokes parameter) in the 
same basis and can be thought of as a $\pi/4$ shear.  
These quantities are related to the original ellipticity
field by a $\varphi$ dependent 
rotation:
\begin{equation}  
\gamma_t=\epp\cos(2\varphi)+\epc\sin(2\varphi); \,\,  
\gamma_{\pi\over 4}=\epc\cos(2\varphi)-\epp\sin(2\varphi).   
\end{equation}  
Figure~2 shows modes where $\gamma_t(\varphi)$ and 
$\gamma_{\pi\over 4}(\varphi)$ 
are independent of $\varphi$.

\begin{figure}
\centerline{\psfig{file=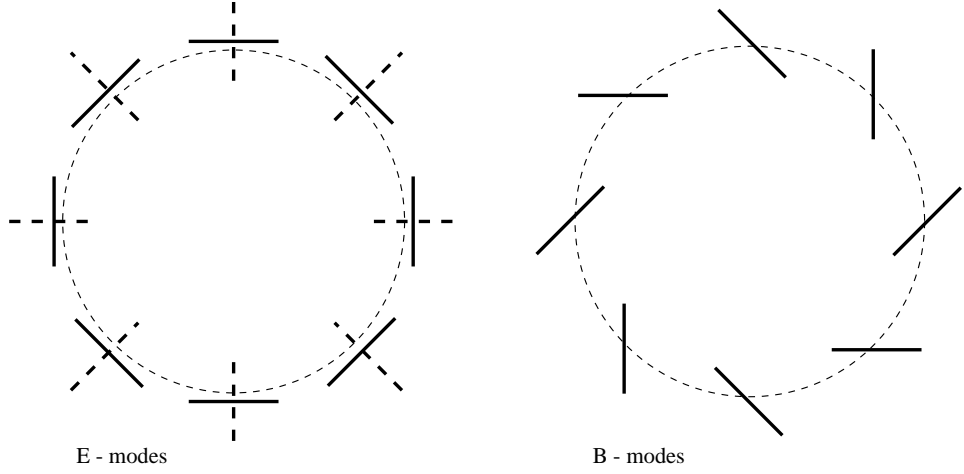,height=2.5in}
}
\caption{Local representations of $E$ (gradient) and $B$ (curl) modes.  
$E$-modes are either tangential or radial, depending on their sign.  
$B$-modes can be oriented in  either a clockwise or counter-clockwise (shown)  
direction.  Lensing generally brings about only E-modes, while noise and 
angular momentum correlations can generate both.   Local estimators 
of the $E$ and $B$-modes can be found by doing a radial 
weighting of these circular integrals (Kaiser et al. 1994).  
}
\label{fig-local}
\end{figure}

It is possible to show that circular integrals of $\gamma_t$ 
and $\gamma_{\pi\over 4}$ are directly related to the $\gamma_E$ 
and $\gamma_B$ contained interior to the circle.  
Using the relations of Eq.~(\ref{eq:defepEB}) in polar coordinates, 
one can show  
\begin{equation}
\gamma_t = {1\over 2} \left( {\partial^2\over\partial r^2}-{1\over
r}{\partial\over\partial r}
-{1\over r^2} {\partial^2\over\partial\varphi^2}\right)\,\Phi_E -
\left({\partial\over\partial r}{1\over
r}{\partial\over\partial\varphi}\right)\,\Phi_B\,;\,\,
\gamma_{\pi\over 4} = {1\over 2} \left( {\partial^2\over\partial r^2}-{1\over
r}{\partial\over\partial r}
-{1\over r^2} {\partial^2\over\partial\varphi^2}\right)\,\Phi_B +
\left({\partial\over\partial r}{1\over
r}{\partial\over\partial\varphi}\right)\,\Phi_E. 
\end{equation}
We can use the polar form of $\nabla^2$ to relate the derivatives of the
potential to $\gamma_E$.  
If we integrate these equations over $\varphi$, the derivatives with respect to 
$\varphi$ drop out and it can be shown that 
\begin{eqnarray}
{1\over 2\pi}\int_0^{2\pi}d\varphi \, \gamma_E ({\bf r})&=& 
{1\over 2\pi}\int_0^{2\pi} d\varphi \, \gamma_t({\bf r}) + 
{1\over \pi r^2}\int_0^r r'dr'\int_0^{2\pi}d\varphi \gamma_E({\bf r'}) 
\nonumber \\
&=& {1\over 2\pi}\int_0^{2\pi} d\varphi \, \gamma_t({\bf r}) + 
2 \int_0^r {dr' \over r'}{1\over 2\pi}\int_0^{2\pi}d\varphi \,\gamma_t({\bf r'})
\end{eqnarray}
A similar relation holds when 
replacing $\gamma_E \rightarrow \gamma_B$ and $\gamma_t  \rightarrow \gamma_{\pi\over 4}$.  

We can convolve these relations with a compensated filter ${\cal{U}}(r)$. 
This filter may be arbitrary, but we will require that 
$\int d^2 r {\cal{U}}(r)=0$.  For example, one might take it to 
have the shape of a Mexican hat. 
Multiplying both sides by $r {\cal{U}}(r)$ and integrating over $r$ 
one obtains the local estimators
\begin{eqnarray}
\Gamma_E \equiv \int d^2r \gamma_E({\bf r}) {\cal{U}}(r) 
&=& \int d^2r \gamma_t({\bf r}){\cal {Q}}(r)  \nonumber \\
\Gamma_B \equiv \int d^2r \gamma_{B}({\bf r}) {\cal{U}}(r)
&=& \int d^2r \gamma_{\pi\over 4}({\bf r}) {\cal {Q}}(r)\,,
\end{eqnarray}
where it follows by integrating by parts that 
${\cal {Q}}(r)={\cal{U}}(r)-{2\over r^2}\int_0^r r' {\cal{U}}(r') dr'$. 
${\cal{U}}(r)$ can be taken to be zero outside a given radius, 
so these relations become purely local.  Thus we have a local measure of the 
$E-B$ decomposition related solely to the ellipticity in that region.  
In the absence of instrumental and sampling noise, 
lensing predicts that $\Gamma_B$ will be identically 
zero for any point on the sky.  

The correlations of these local measures can be computed as follows,
\begin{eqnarray}
\langle\Gamma_E(0)\Gamma_E(R)\rangle 
&=& \int {d^2 k\over 2\pi}  \hat{{\cal{U}}}^2(k)
e^{i{\bf k}\cdot{\bf R}} \int d^2r \langle\gamma_E(0)\gamma_E(r)\rangle e^{i{\bf k}\cdot{\bf r}}\nonumber\\
&=& \int d^2r {1\over 2}\left[(\xip(r)+\xic(r))+\nabla^4\chi^{-1}\left\{\xip(r)-\xic(r)\right\}\right]
{\cal {W}}(|{\bf r+R}|)\nonumber\\
&=& {1 \over 2}\int d^2r [\xip(r)+\xic(r)] 
{\cal {W}}(|{\bf r+R}|) + {1\over 2}\int d^2r
[\xip(r)-\xic(r)] \tilde{\cal {W}}(|{\bf r+R}|)\,,
\end{eqnarray}
where we used Eq.~(\ref{eq:invE}), and defined $\hat{\cal {W}}(k)\equiv
\hat{\cal{U}}^2(k)$, 
so that ${\cal {W}}(r)$ is the convolution of ${\cal{U}}(r)$ with itself.  
The function $\tilde{\cal {W}}(r)$ can be shown to be a convolution of 
${\cal {W}}(r)$ 
and ${\cal{G}}(r,r')$, i.e.  
\begin{equation}
\int r dr {\cal {W}}(r)\nabla^4\chi^{-1} g(r) 
= \int rdr {\cal {W}}(r) \int r'dr' {\cal{G}}(r,r')
g(r') = \int r'dr' g(r') \tilde{\cal {W}}(r')\,.
\end{equation}
From this it follows that 
$\tilde{\cal {W}}(r')\equiv\int rdr {\cal {W}}(r) {\cal{G}}(r,r') 
= \chi\nabla^{-4}{\cal {W}}(r')$. 
The corresponding expression for
the $B$-mode is
\begin{equation}
\langle\Gamma_B(0)\Gamma_B(R)\rangle ={1\over 2}\int d^2r 
[\xip(r)+\xic(r)] {\cal {W}}(|{\bf r+R}|) - {1\over 2}\int d^2r
[\xip(r)-\xic(r)] \tilde{\cal {W}}(|{\bf r+R}|)\,.
\end{equation}
The variances of the $E$ or $B$ field smoothed with 
a given window ${\cal{U}}(r)$ 
is obtained by setting the separation $R=0$. 
Note that as in the previous section, 
we are implicitly incorporating statistical 
isotropy in these correlation expressions.  

For concreteness, it is useful to consider a simple example of a wavelet shape.
Following work by van Waerbeke (1998). 
assume the radial function to have the form 
of a Mexican hat wavelet, which is the 
derivative of a
Gaussian function, ${\cal{U}}(r)=\sigma^{-2}\,(1-r^2/
2\sigma^2)\,\exp(-r^2/2\sigma^2)$ and its Fourier transform is simply 
$\hat{{\cal{U}}}(k) = {1\over 2}k^2\sigma^2 e^{-k^2\sigma^2/2}.$  
For this particular choice, the convolution of ${\cal{U}}(r)$ with itself is 
${\cal {W}}(r) = (1/2\sigma^2)\left[2-r^2/\sigma^2+r^4/16\sigma^4\right]\,
\exp(-r^2/4\sigma^2)$. 
Finally, using the fact that the
Fourier representation of $\nabla^{-4}$ is $k^{-4}$, we have,
\begin{equation}
\tilde{{\cal {W}}}(r)= \chi\left[{\sigma^2\over2}\,e^{-r^2/4\sigma^2}\right] =
{1\over 2\sigma^2}\,\left({r^2\over4
\sigma^2}\right)^2\,e^{-r^2/4\sigma^2}\,.
\end{equation}
This particular wavelet has the advantages that it is simple, analytic and 
very compact, falling off exponentially at large distances.  

\section{Data Analysis} 
In this section we address the direct analysis of real survey data.  We
first consider the extraction of the $E$ and $B$ correlators.  For this
purpose, one only needs to measure the two pairwise ellipticity
correlation functions $\xi_+,\xi_\times$ (defined before Eqn (15)) for
all pairs of galaxies as a function of separation, which has been done
by van Waerbeke et al, (2000) and does not depend on the geometry of the
survey or its boundary shapes.  This correlation function (plus its
error bars and the covariance matrix of errors) contains all the second
order statistics of the map, and is a complete and optimal two point
description.  

We now have two functions, both of which contain
noise, and our goal is to apply a rotation such that one function
contains lensing signal and the other no lensing signal, but which will give 
an estimate of contamination from noise and intrinsic correlations. 
The current
analyses effective have added the two correlations, which adds a
function which contains lensing to one which contains no lensing but
an equal amount of noise, which doubles the amount of noise we have.
Instead, we can define 
$\xi'(r)=2/r^2 \int_0^r [\xi_+(r')+ \xi_\times(r')] r'dr' -6/r^4
\int_0^r [\xi_+(r')+ \xi_\times(r')] r'^3dr'$.  
We can now derive pure $E$-type and $B$-type correlators which depend only on 
correlations at separations less than $r$, 
\begin{eqnarray}
 \chi\,\nabla^{-4} \xi_E(r) &=& \int k dk J_4(kr) P_E(k) = 
\xi_+(r)+\xi'(r) \nonumber \\
 \chi\,\nabla^{-4} \xi_B(r) &=& \int k dk J_4(kr) P_B(k) = 
\xi_\times(r)+\xi'(r) 
\label{eqn:dummy}
\end{eqnarray}
In the case of pure weak lensing, the $B$-type correlator should be
consistent with pure noise, while $\xi_E$ contains all the lensing
signal, and only half the noise.  We have achieved the correlation function
analogy of Kaiser's 45 degree rotation: rotating all the images by
45 degrees swaps $\xi_E$ and $\xi_B$.  

One can also obtain expressions for the variances of the fields smoothed by a 
tophat filter with radius $R$.  
This is done by convolving the correlation functions with a window which is 
proportional to the area of overlap between two circles of radius $R$ and 
separation $r$, 
\begin{equation}
\langle \gamma_E^2(R)\rangle_{TH} =\frac{2}{\pi R^4}
  \int_0^{2R} r dr \xi_E(r) \left(2 R^2\cos^{-1}(r/2R)-r\sqrt{R^2-r^2/4}\right).
\label{eqn:TH}
\end{equation}
This again contains half as much noise power as the standard procedure
of actually convolving the map and computing its variance, which is
what all work to date has performed.

This decomposition was performed directly on the correlator, and there
exists no transformation on the shear map such that the correlation
function of the transformed shear map is given by Eqn. (\ref{eqn:dummy}).
Similarly, Eqn. (\ref{eqn:TH}) is not identically equal to the tophat
smoothed map variance.  If one actually wants to make a map, the aperture
shear decomposition Eqn. (28) allows one to make a local decomposed map,
whose variance one can then measure.  Note that in making a map, one looses
information near the boundaries, and variation in source counts leads to
inhomogeneous signal-to-noise, decreasing the overall sensitivity to
measuring the true correlation function.  Luckily the aperture mass
approach also allows a direct computation of the same quantities from
the correlation functions, as shown in Eqn. (31).

\section{Conclusions} 

Here we have investigated the general decomposition of flat two dimensional 
spin-2 fields into so-called electric (gradient) and magnetic (curl) 
components. While this decomposition involves derivatives and is 
intrinsically non-local, we have shown how local correlations of the electric 
and magnetic components can be found given correlations in the 
components of the 
ellipticity  (Equations 17 and 18).  
For the case of power law correlations, this 
implies a relationship between the spectral index of the correlations 
and the relative amplitude of the different types of ellipticity correlations. 

In addition, following Kaiser et al. (1994) 
we have shown how local estimators for the electric 
and magnetic modes can be constructed from circular integrals
of the tangential and $\pi/4$ (or $Q$ and $U$ Stokes parameters) 
distortions respectively (Figure 2). 
We calculated correlations of these local estimators and related them to 
the  ellipticity correlations.  

This decomposition has important consequences when applied to the 
projected shapes of galaxies.  Gravitational lensing primarily produces 
electric modes, as does the tidal stretching of galaxies.
However, as we have shown here, angular momentum couplings 
produce $E$ and $B$-modes in comparable amounts and 
one might expect that noise, 
telescope distortions and other sources of systematic errors will 
produce curl modes as well. 

Thus the presence of $B$-modes will be useful for disentangling intrinsic 
correlations caused by angular momentum couplings from those 
induced by cosmic shear and from gravitational lensing.  
Even if lensing distortions dominate,  
this decomposition will be useful as a means of estimating the levels of noise
and systematic errors in the observations.  In addition, it provides a means 
of reducing noise levels of lensing observations by a factor of $\sqrt{2}$ 
(Kaiser 1995.) 

The prospects for isolating the contribution of $B$-modes using
these local correlators are promising
with several of the ongoing shallow redshift and imaging
surveys presently taking data. 
These include 
the Sloan DSS ({\tt www.sdss.org}), 2dF ({\tt www.ast.cam.ac.uk/AAO/2df}), 
2MASS ({\tt pegasus.phast.umass.edu}),  DEEP ({\tt dls.bell-labs.com}) and 
the INT wide field survey ({\tt www.ast.cam.ac.uk/$\sim$wfcsur/}).
The low median
redshift implies minimal contamination from the lensing signal
and therefore an improvement in the signal to noise of the
extraction.

Finally, most of these considerations apply equally well to the imminent 
observations of CMB polarization 
(e.g. Kamionkowski, Kosowsky \& Stebbins 1997; Zaldarriaga \& Seljak 1997).    
In this case, scalar 
fluctuations induce only $E$-modes, while noise and gravitational radiation 
induce both $E$ and $B$-modes.  
As in the case of galaxy shapes, these observations
will be initially noise dominated, so these kinds of 
correlation analyses will be essential.   The lessons we learn from the 
lensing data now available will be directly applicable to the polarization 
data when they become available in a few years time.   
 
\begin{acknowledgements}
We thank Ben Metcalf, 
Neil Turok and Ludo van Waerbeke for useful conversations.  
RC and TT acknowledge PPARC for the award of an Advanced and a
post-doctoral fellowship, respectively.
\end{acknowledgements}

\vfill\eject
\end{document}